\documentclass[sigconf]{acmart}
\usepackage[utf8]{inputenc}
\usepackage{booktabs} 
\usepackage{multirow}
\usepackage{graphicx}
\usepackage{array}
\newcolumntype{C}[1]{>{\centering}m{#1}}
\newcolumntype{L}[1]{>{\raggedright\let\newline\\\arraybackslash\hspace{0pt}}m{#1}}

\usepackage{footnote}
\makesavenoteenv{tabular}
\makesavenoteenv{table}

\usepackage{color}
\usepackage{colorspace}
\definespotcolor{mygreen}{PANTONE 3536 C}[rgb]{0, 114, 0.14}

\usepackage{adjustbox}

\usepackage{balance}

\copyrightyear{2020}
\acmYear{2020}
\setcopyright{acmlicensed}
\acmConference[ddh20]{ddh20: Data and Digital Humanities 2020}{October 15--17, 2020}{Hammamet, Tunisia}
\acmBooktitle{ddh20: Data and Digital Humanities 2020, October 15--17, 2020, Hammamet, Tunisia}
\acmPrice{}
\acmDOI{}
\acmISBN{}

\begin{document}
	\title{Data Lakes for Digital Humanities}

\author{Jérôme Darmont}
\email{jerome.darmont@univ-lyon2.fr}
\orcid{0000-0003-1491-384X}
\author{Cécile Favre}
\email{cecile.favre@univ-lyon2.fr}
\orcid{0000-0002-8658-7564}
\author{Sabine Loudcher}
\email{sabine.loudcher@univ-lyon2.fr}
\orcid{0000-0002-0494-0169}
\affiliation{
  \institution{Université de Lyon, Lyon 2, ERIC~UR~3083}
  \streetaddress{5 avenue Pierre Mendès France}
  \city{Bron}
  \postcode{F69676}
  \country{France}
}

\author{Camille Noûs}
\email{camille.nous@cogitamus.fr}
\orcid{0000-0002-0778-8115}
\affiliation{
  \institution{Université de Lyon, Lyon 2, Laboratoire Cogitamus}
  \streetaddress{5 avenue Pierre Mendès France}
  \city{Bron}
  \postcode{F69676}
  \country{France}
  }
  
	\begin{abstract}
Traditional data in Digital Humanities projects bear various formats (structured, semi-structured, textual) and need substantial transformations (encoding and tagging, stemming, lemmatization, etc.) to be managed and analyzed. To fully master this process, we propose the use of data lakes as a solution to data siloing and big data variety problems. We describe data lake projects we currently run in close collaboration with researchers in humanities and social sciences and discuss the lessons learned running these projects.
	\end{abstract}
	
	%
	%



\renewcommand{\shortauthors}{Darmont \& al.}
	
\maketitle
	
\section{Introduction}
\label{sec:intro}

Traditional data management has long been adopted by many researchers involved in Digital Humanities (DH). However, it requires a substantial investment in data modeling, including, at the physical level, technologies such as relational and semi-structured Database Management Systems (DBMSs), various data formats, e.g., XML and JSON for semi-structured data, RDF for linked data, and query languages such as SQL and XQuery. This investment in computer science and the fact that initial data are inevitably transformed are presumably impediments to the adoption of DBMSs and related digital tools for DH.

Moreover, most source information exploited by humanities and social sciences comes in textual format. Again, such textual documents are difficult to manage without substantial transformations: digitization, encoding and tagging, e.g., via the Text Encoding Initiative (TEI), and even lowercasing, stemming, lemmatization, stopword removal or normalization when it comes to text mining and natural language processing.

Another important methodological issue is the black box effect that occurs when resorting to computer scientists only ``as a service''. How can DH researchers work without mastering the whole process? Furthermore, designing and managing such processes also  lead to research issues for computer scientists.

To leverage the above-mentioned issues, we propose the use of data lakes, a concept introduced by Dixon in 2010 as a solution to data siloing and big data variety problems \cite{Dixon2010}. Even if data exploited by DH are not always big data in terms of volume, they can bear considerable variety, i.e., including structured and semi-structured data, as well as unstructured data such as texts, various types of images, sounds and videos. Traditional data management tends to manage such heterogeneity with different systems, thus separating data into so-called silos.

A data lake is a scalable storage and analysis system for data of any type, retained in their \emph{native format} and used \emph{mainly} (but \emph{not only}) by data specialists (statisticians, data scientists or analysts) for knowledge extraction \cite{bbigap19}. 

One of the main advantages of data lakes is that data are stored in their initial form, and are thus recognizable by their producers, such as DH researchers. A data lake does not propose a new data model nor new data formats for data archiving. Moreover, when data are transformed for processing, the data lineage is stored as metadata, thus enforcing traceability. 

However, a drawback is that unprepared data are difficult to process and require data specialists who can program. Yet, we strongly advocate, with other researchers, for the ``industrialization'' of data lakes, i.e., providing a software layer that allows non-data scientists such as DH researchers to transform and analyze their own data in autonomy, just as dynamic reports are prepared on top of data warehouses for the use of business (i.e, non technical) users.

The remainder of this paper is organized as follows. In Section~\ref{sec:projects}, we describe data lake projects we currently run in close collaboration with researchers in social sciences and humanities. In Section~\ref{sec:conclusion}, we conclude this paper by discussing the lessons learned running these projects.

\section{Example DH projects involving data lakes}
\label{sec:projects}

\subsection{HyperThesau}
\label{sec:hyperthesau}

The ``Hyper thesaurus and data lakes: Mine the city and its archaeological archives'' (HyperThesau) project involves a multidisciplinary team consisting of two research laboratories of archaeology and of computer science, a digital library, two archaeological museums and a private company. This project has two main objectives: 
\begin{enumerate}
    \item the design and implementation of an integrated platform to host, search, share and analyze archaeological data;
    \item the design of a domain-specific thesaurus taking the whole archaeological data lifecycle into account, from data creation to publication.
\end{enumerate}

Archaeological data may bear many different types, e.g., textual documents, images (photographs, drawings...), sensor data, etc. Moreover, similar documents, e.g., excavation reports, are often created by various software tools that are not compatible with each other. The description of an archaeological object also differs with respect to users, usages and time. Such a variety of archaeological data induces many scientific challenges related to storing heterogeneous data in a centralized repository, guaranteeing data quality, cleaning and transforming the data to make them interoperable, finding and accessing data efficiently and cross-analyzing the data with respect to their spatial and temporal dimensions. 

\begin{figure}[hbt]
    \centering
    \includegraphics[width=8.5cm]{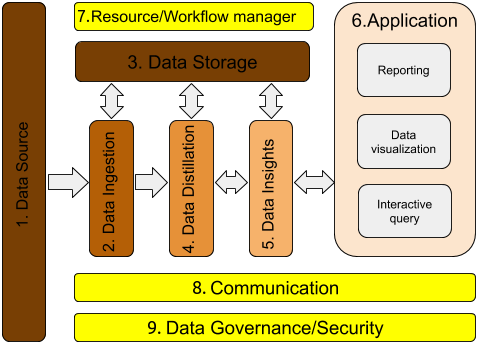}
    \caption{HyperThesau data lake's layered architecture \cite{Liu2020}}
    \label{fig:HTarchi}
\end{figure}

To overcome all these challenges, we implement a data lake. Our approach aims to collect all types of archaeological data, save them inside the data lake and propose metadata for better organizing data and for allowing users to easily find data for analysis purposes. 
Our data lake prototype is architectured in nine layers (Figure~\ref{fig:HTarchi}) \cite{dh20,Liu2020}. 

\begin{enumerate}
    \item The data source layer gathers the basic properties of data sources, e.g., volume, format, velocity, connectivity, etc. Based on these properties, data engineers can determine how to import data into the lake.
    \item The data ingestion layer provides a set of tools for performing batch or real-time data integration. Data engineers can choose the right tools and plans to ingest data into the data lake with respect to data source properties and the lake's capacity. During ingestion, metadata  provided by data sources, e.g., name of excavation sites or instruments, must be gathered as much as possible.
    \item The data storage layer is core to a data lake. It must have the capacity to store all data in any format.
    \item The data distillation layer provides a set of tools for data cleaning (eliminating errors such as duplicates and type violations) and encoding formalization (converting various data and character encoding). 
    \item The data insights layer provides a set of tools for data transformation (e.g., into models) and exploratory data analysis (e.g., pattern discovery). Note that transformed data may also be stored into the lake for later reuse.
    \item The data application layer provides applications that allow users extracting value from data, e.g., through an interactive query system, reports or dataviz.
    \item The workflow manager layer provides tools to automate the flow of data processes.  
    \item The communication layer provides tools that allow the other layers to communicate with each other. It must provide synchronous and asynchronous communication capability. 
    \item The data governance layer provides a set of tools to establish and execute plans and programs for data quality control \cite{Vijay2010}.
\end{enumerate}

Each of the above layers is implemented with one or more frameworks of the Apache Hadoop ecosystem, e.g., Atlas\footnote{\url{https://atlas.apache.org}}, HDFS\footnote{\url{hhttps://hadoop.apache.org/docs/r1.2.1/hdfs_user_guide.html}}, HIVE\footnote{\url{https://hive.apache.org}}, OpenLdap\footnote{\url{https://www.openldap.org}}, Spark\footnote{\url{http://spark.apache.org}}, etc. This prototype is operational and currently hosts the data of two archaeological research facilities. The metadata management system instantiates the MEtadata model for Data Lakes (MEDAL), which adopts a graph model \cite{bbigap19}. It is implemented with Apache Atlas, which can host not only descriptive metadata, but also several thesauruses. With the help of a search engine, i.e., Solr\footnote{\url{https://lucene.apache.org/solr/}}, users can find data through descriptive metadata, a thesaurus or the data lineage.

\subsection{Bretez/STRATEGE}
\label{sec:bretez}

Bretez \cite{bretez} is a multidisciplinary project aiming at a visual and sonorous restitution of the XVIII\textsuperscript{th}-century Paris. It is also an exploratory project constituted of successive, interlinked modules that are (and must be) interoperable and open. The historical urban restitution is achieved with video game engines that bear their own respective characteristics, of course related to gaming. Yet, here, they are used for specific management and traceability needs. Moreover, Bretez' documentation is a voluminous corpus of heterogeneous and multimedia data.

\begin{figure*}[hbt]
    \centering
    \includegraphics[width=17.5cm]{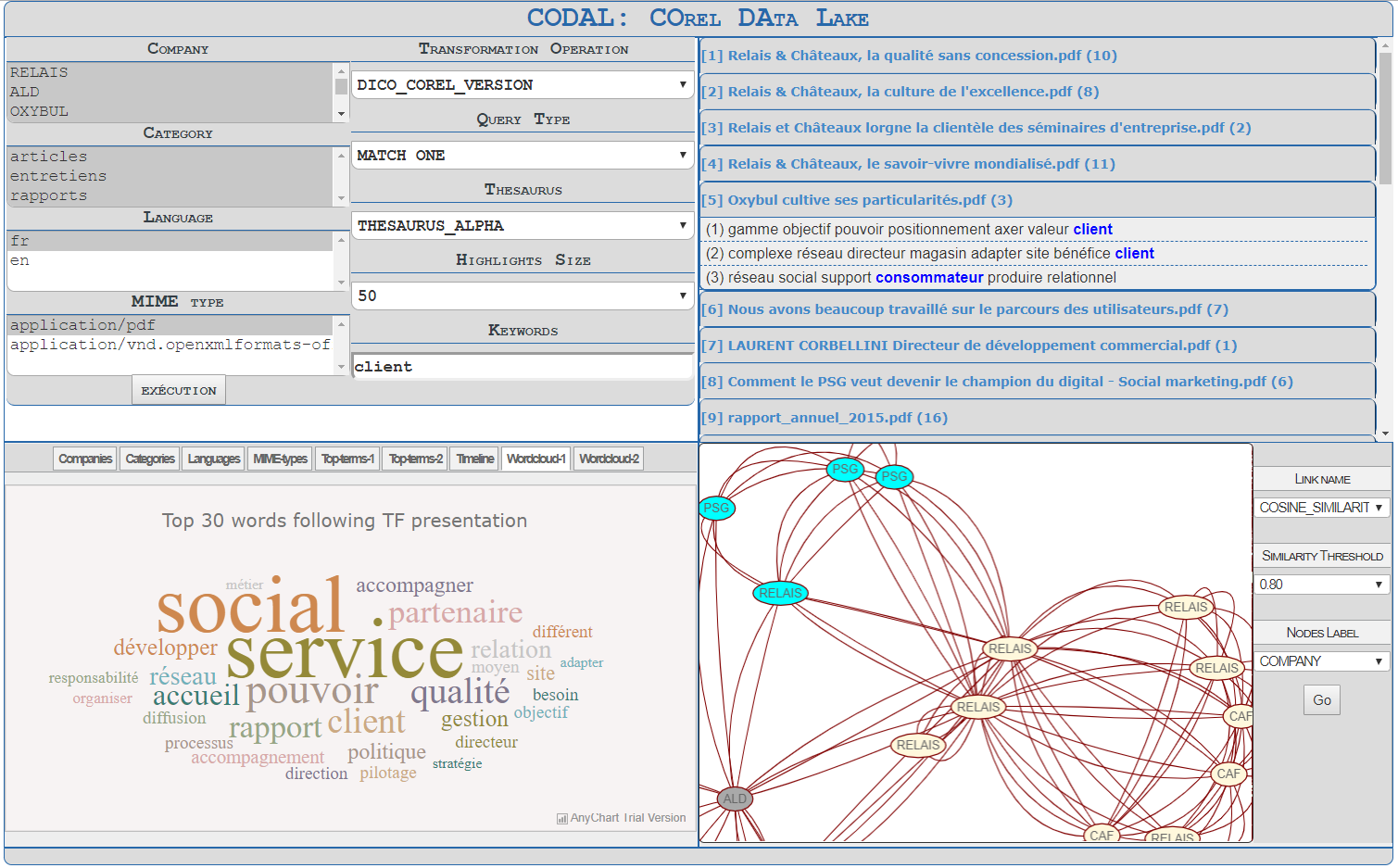}
    \caption{CODAL example screenshot \cite{iceis19}}
    \label{fig:CODAL}
\end{figure*}

Within project Bretez, the ``Traceability and information management system for multimedia data'' (STRATEGE) aims at designing, storing, querying and analyzing all the project's data. To master data heterogeneity, manage data quality and volume, warrant data interoperability and an efficient access while keeping data in their original form so that they remain usable reference for the project's researchers, we resort to a data lake.

STRATEGE is in its first stages: we catalogued all existing data, which included a database, textual documents, sounds, images and a 3D Unity\footnote{\url{https://unity.com}} (a game engine) model, and more. The database is particularly interesting, for it contains both data and metadata.
While retaining it, we also restructured it so as to allow its metadata to interoperate with specific data lake metadata. In short, there are ``business'' metadata and technical metadata.

The remaining tasks include fully designing and integrating the metadata system, on the basis of MEDAL \cite{bbigap19}; make the data from the Unity model accessible into the lake; formalize analysis needs; and design tools that must jointly handle textual, visual and audio content, as well as the heterogeneity of data sources. Such software tools must be accessible to all researchers involved in the Bretez project.

\subsection{COREL and AURA PMI}
\label{sec:corel}

Both the projects ``At the heart of customer relationship'' (COREL) and ``Digital transformation, servicization and mutations of industrial SME business models'' (AURA PMI) relate to management sciences and are carried out in collaboration with the Coactis laboratory\footnote{\url{http://coactis.org}}. Although their respective focus and scope are different, they are quite similar in terms of data: a corpus of various textual documents (e.g., annual reports from companies and organizations, interviews of senior or top executives, press articles; all in French or English) and data from various sources, including the Web, curated and inferred by researchers in management sciences from company legal information and performance indicators such as workforce, annual revenue, stock-exchange price and perceived level of digitization and servicization.

With such data handy, the objective is to cross-analyze the terms and expressions found in textual resources with structured, qualitative and quantitative data, in order to discover new insights regarding how companies communicate \textit{vs.} their actual customer relationship management strategy (for COREL) and how digitization and servicization impact economic performance (for AURA PMI). The challenges here are to:
\begin{enumerate}
    \item leverage metadata that allow querying the whole corpus; 
    \item jointly analyze structured and unstructured data; 
    \item allow management science researchers to perform analyses by themselves.
\end{enumerate}

To complete these tasks for the COREL project, we designed a metadata system that prefigured MEDAL \cite{bbigap19} and proposed the lightweight COREL Data Lake architecture (CODAL) \cite{iceis19}, which is composed of:
\begin{enumerate}
    \item a storage layer that notably includes Elasticsearch\footnote{\url{https://www.elastic.co/elasticsearch/}} 
for indexing textual contents;
    \item a metadata layer leveraging and extending the Metadata Encoding \&  Transmission Standard (METS) \cite{TLoC2017}, stored in the BaseX XML DBMS\footnote{\url{http://basex.org}};
    \item an analysis layer, i.e., an intuitive Web-based graphical interface that allows management science researchers to perform analyses in autonomy, thus enforcing the ``industrialization'' of CODAL.
\end{enumerate}

The analysis layer features three kinds of analyses:
\begin{enumerate}
    \item data exploration akin to On-Line Analytical Processing (OLAP)\cite{olap93};
    \item proximity analyses such as similarity (what  documents  are  similar  or  different \cite{Pons2006}) and centrality (to identify the documents bearing a specific or common vocabulary, hinting at its importance \cite{Farrugia2016}) analyses;
    \item custom highlights of the context of terms and, optionally, their synonyms, in textual documents.
\end{enumerate}
All three types of analyses come with various dataviz (Figure~\ref{fig:CODAL}).

The AURA PMI Data Lake (AUDAL) is currently being developed, and builds upon CODAL. Its metadata system will notably be a substantial evolution of MEDAL supported by the Neo4J\footnote{\url{https://neo4j.com}} graph DBMS. Moreover, the AUDAL analysis layer, which lays on an Application Programming Interface (API), will be much more elaborate and efficient than CODAL's.


\section{Conclusion}
\label{sec:conclusion}

In all four data lake projects summarized in Section~\ref{sec:projects}, we use different versions of the MEDAL metadata system, which is designed to be generic. However, although MEDAL is quite flexible, we do not believe in a single model for data lakes. There are indeed significant differences in data in only four projects, in terms of volume, variety and velocity, which imply different architectures and technologies. Thus, we think that much needed methodological tools for data lakes should be \textit{instantiated} for each project rather than applied ``as is''.

Furthermore, the software layer we add to ``industrialize'' our data lakes might become yet another black box, while there is a strong stake for researchers in humanities and social sciences involved in DH projects not to be dispossessed of data by an analysis layer that would adopt a ``click and go'' approach. Data are indeed often partly constructed by said researchers themselves as a product of scientific work that takes time, thus giving a significant value to datasets.

In consequence, we take great care of accompanying DH users in their appropriation of our analysis tools, not only by training, but especially by interweaving research methodologies from computer science and other disciplines \textit{by design}, in close collaboration with partner researchers.

Moreover, the possibility of having both access to the raw data and the entire possible processing chain is necessary, because black boxes are seldom compatible with a sound methodological approach aiming at producing scientific knowledge. Data lakes precisely allow this much needed transparency.

\section*{ACKNOWLEDGEMENTS}

Projects HyperThesau and Bretez/STRATEGE are funded by the Laboratory of Excellence ``Intelligence of Urban Worlds'' (IMU)\footnote{\url{https://imu.universite-lyon.fr}}. Project COREL was funded by the University of Lyon~2. Project AURA PMI is funded by the Auvergne-Rhône-Alpes Region.

\balance
	\bibliographystyle{ACM-Reference-Format}
	\bibliography{dl4dh.bib}

\end{document}